\documentclass[onecollarge]{svjour2}       
\usepackage{graphicx}
\def\geqsim{\mathbin{\;\raise1pt\hbox{$>$}\kern-8pt\lower3pt\hbox{$\sim$}\;}}
\def\leqsim{\mathbin{\;\raise1pt\hbox{$<$}\kern-8pt\lower3pt\hbox{$\sim$}\;}}

\begin{document}

\title{Asymptotically charged BTZ Black Holes in Gravity's Rainbow}
\author{S. H. Hendi$^{1,2}$\thanks{\emph{Present address:} hendi@shirazu.ac.ir}} \institute{$^1$Physics
Department and Biruni Observatory, College of Sciences, Shiraz
University, Shiraz 71454, Iran\\
$^2$Research Institute for Astrophysics and Astronomy of Maragha
(RIAAM), P.O. Box 55134-441, Maragha, Iran}

\date{Received:  / Accepted: }

\maketitle

\begin{abstract}
Motivated by the wide applications of BTZ black holes and
interesting results of gravity's rainbow, we consider three
dimensional rainbow solutions and investigate their thermodynamic
properties. In addition to investigate black holes thermodynamics
related to AdS/CFT correspondence, one may regard gravity's
rainbow to to encode quantum gravity effects into the black hole
solutions. We take into account the various models of linear and
nonlinear electrodynamics and study their effects on the gravity's
rainbow spacetime. We also examine thermal stability and find that
obtained three dimensional rainbow black holes are thermally
stable.
\end{abstract}

\section{Introduction\label{Intro}}

One of the weighty dream of physicists is creating a consistent
quantum theory of gravity. Although various aspects of the quantum
gravity is still very much alive, till now, all the attempts of
obtaining a complete description of the quantum gravity have been
unfeasible. Gravity's rainbow arises from deep attempts to fix the
gaps between the theory of general relativity and quantum
mechanics. One of the possible interesting consequence of
gravity's rainbow is that the universe had no beginning, and
therefore time is stretched back infinitely without encountering a
big bang \cite{Awad-Ali}. Also, thermodynamical properties of
black p-branes was reported within the framework of gravity's
rainbow \cite{pBrane}. Furthermore, as more recent application of
gravity's rainbow, a recent investigation on the possibility of
resolving the black hole information paradox was carried out in
Ref. \cite{paradox}. In addition, the effects of the gravity's
rainbow on the deflection of light, photon time delay,
gravitational red-shift, and the weak equivalence principle have
been studied in Ref. \cite{Solar}. Other consequences of the
gravity's rainbow in the context of the gravitational collapse and
black hole physics have been investigated in literature (see
\cite{GravRain1,GravRain2,GravRain3,GravRain4} and references
therein).

On the other hand, some of theoreticians believe that the violation of
Lorentz invariancy is an essential rule to construct quantum theory of
gravity. The Lorentz invariance violation may be expressed in form of
modified dispersion relations \cite%
{Kostelecky,Gambini,Carroll,Camelia,tHooft}. In other words, according to
the loop quantum gravity (LQG) results \cite{Gambini} and spacetime
discreteness assumption \cite{tHooft} at the Planck scale, one may regard
Lorentz invariance violation or modified dispersion relation by redefining
the physical momentum and energy. In order to have a deep insight into the
Lorentz invariancy, one may regard the close relation between gravity's
rainbow and Horava-Lifshitz gravity \cite{Garattini2015}. This relation
comes from the fact that both these theories are based on the breaking of
the usual energy-momentum dispersion relation in the UV limit in such a way
that it reduces to the usual energy-momentum dispersion relation in the IR
limit. The UV modification of the usual energy-momentum relation implies the
breaking of the Lorentz symmetry. Due to the existence of an unstable
perturbative string vacuum, the spontaneous breaking of the Lorentz symmetry
can be occurred in string theory \cite{Kostelecky}. In addition, the Lorentz
symmetry breaking can be studied using black brane in the context of type
IIB string theory \cite{Mukohyama2007}. Taking into account the mentioned
motivation of UV deformation of geometries that occur in string theory and
the close relation between Horava-Lifshitz gravity and gravity's rainbow
\cite{Garattini2015}, recently rainbow deformation of geometries has been
performed, in which leads to modification of the dispersion relation. One of
the modified dispersion relation, which is predicted by new approach of
quantum gravity \cite{Horava1,Horava2}, takes the following general form
\begin{equation}
f^{2}{(\varepsilon )}{E}^{2}-g^{2}{(\varepsilon )}{p}^{2}=m^{2},  \label{MDR}
\end{equation}%
where dimensionless relative energy ${\varepsilon =E/{{E}_{p}}}$,
in which $E $ is the energy of the probing particle and $E_{p}$ is
the Planck energy. In addition, temporal and spatial rainbow
functions ($f({\varepsilon })$ and $g({\varepsilon })$) satisfy
the following conditions
\begin{equation}
\lim\limits_{{\varepsilon }\rightarrow 0}f({\varepsilon })=1,\qquad
\lim\limits_{{\varepsilon }\rightarrow 0}g({\varepsilon })=1.
\end{equation}

On the other hand, such modified dispersion relation can be
originated from the so-called doubly special relativity
\cite{DSR}. Doubly special relativity is a natural extension of
special relativity, which enjoys the invariancy of the speed of
light, to the case of assuming the invariancy of the maximum
energy scale (the Planck energy) or equivalently minimum length
scale \cite{Magueijo2002}. Besides, the generalized uncertainty
principle (GUP) which comes from various approaches to quantum
gravity, is based on the existence of minimum measurable length
scale \cite{GUP1,GUP2,GUP3,GUP4,GUP5,GUP6,GUP7}. Such minimum
length scale has been applied to black hole thermodynamics
\cite{GUPThermo1,GUPThermo2,GUPThermo3,GUPThermo4,GUPThermo5}.
From string theory point of view, it is not possible to probe
spacetime below the string length scale and one can translate the
minimum length scale into a maximum energy scale
\cite{GUPenergy1,GUPenergy2,StringRainbow1,StringRainbow2,StringRainbow3,StringRainbow4}.
Including the curvature into the doubly special relativity,
Magueijo and Smolin proposed the so-called doubly general
relativity \cite{Smolin}. In other words, doubly general
relativity is a natural extension of doubly special relativity to
the case of assuming curved spacetime. Regarding the doubly
general relativity, one finds the geometry of spacetime may depend
on the energy of the particle moving in it. In other words,
spacetime is parameterized by the ratio ${\varepsilon =}E/E_{P}$
to obtain a parametric family of metrics, the so-called rainbow of
metrics \cite{Smolin}. Hence, the modified metric in gravity's
rainbow can be written as \cite{Smolin}
\begin{equation}
g({\varepsilon })=\eta ^{ab}e_{a}({\varepsilon })\otimes e_{b}({\varepsilon }%
),  \label{Rmetric}
\end{equation}%
in which the energy dependence of the frame fields is
\begin{equation}
e_{0}({\varepsilon })=\frac{1}{f({\varepsilon })}\hat{e}_{0},\qquad e_{i}({%
\varepsilon })=\frac{1}{g({\varepsilon })}\hat{e}_{i},
\end{equation}%
where the hatted quantities refer to the energy independent frame
fields. In Ref. \cite{GravRain4} it was shown that one can explain
the absence of black holes at LHC by using the gravity's rainbow.
In addition, regarding gravity's rainbow, one may resolve the
black hole information paradox without any need for black hole
complementarity or Firewall \cite{paradox}. The functional forms
of $f({\varepsilon })$ and $g({\varepsilon })$ are based on
various
phenomenological motivations \cite%
{Awad-Ali,paradox,Garattini2012,Leiva2009,Li-Ling2009,AliPRD2014,Barrow,Liu-Zhu,HendiFaizal}%
. Among different proposals in the literature, we point out three
interesting models of modified dispersion relation.

One of the most studied models was proposed in Refs. \cite%
{Amelino1996,amelino2013}. This model, which we identify it as
first model, is compatible with the results of LQG and
non-commutative spacetime \cite{amelino2013}. In this model one
should set the temporal rainbow's function to unity and spatial
one is a square root function with the following explicit forms
\begin{equation}
f\left( {\varepsilon }\right) =1,\quad g\left( {\varepsilon }\right) =\sqrt{%
1-\eta {\varepsilon }^{n}},
\end{equation}

Amelino-Camelia, et al. proposed another model in Ref. \cite{Camelia}, which
we will call second model. The hard spectra of gamma-ray bursters may be
explained by the second model \cite{Camelia}. In this model one sets spatial
rainbow function to unity and temporal one has exponential form
\begin{equation}
f\left( {\varepsilon }\right) =\frac{{{e}^{\sigma {\varepsilon }}}-1}{\sigma
{\varepsilon }},\quad g\left( {\varepsilon }\right) =1,
\end{equation}

Another model, which we will recognize it as third model, is related to
constant speed of light and one may use it to solve the horizon problem \cite%
{Smolin}. In this model both rainbow's functions are the same with the
following fractional form
\begin{equation}
f\left( {\varepsilon }\right) =g\left( {\varepsilon }\right) =\frac{1}{%
1-\lambda {\varepsilon }}.
\end{equation}

Regarding the three mentioned models, it is notable that the
dimensionless parameters $\eta $ , $n$, $\sigma $ and $\lambda $
are of the order of unity.

Among all black hole solutions one can think of, BTZ (Banados-
Teitelboim-Zanelli) \cite{BTZ1,BTZ2} black hole can be considered
as the unique one. It is believed that $(2+1)-$dimensional BTZ
black hole solutions have great achievements and provide a
simplified model to find some conceptual issues in the context of
black hole physics, quantum gravity, string and gauge theory, and
also in the context of the AdS/CFT correspondence
\cite{BTZapp1,BTZapp2,BTZapp3,BTZapp4,BTZapp5,BTZapp6,BTZapp7}. In
addition, critical behavior of the BTZ black hole with torsion has
been considered in Ref. \cite{BTZtorsion}. Also, Gregory-Laflamme
instability of BTZ black hole in the context of massive gravity
has been reported by Moon and Myung \cite{GLstability}.
Thermodynamic equilibrium states of a typical static thin shell in
a $(2+1)-$dimensional spacetime have been analyzed in
\cite{ThinShell} and thermodynamical properties of a BTZ black
hole solution with an Horndeski source have been investigated in
Ref. \cite{BTZHorndeski}. For more subjects related to BTZ
solutions, we refer the reader to Refs.
\cite{BTZa,BTZb,BTZc,BTZd,BTZe,BTZf,BTZg,BTZh,BTZi,BTZj,BTZk,BTZl}
and references therein.

Recently, the effective potential structure of the neutral BTZ
black hole in gravity's rainbow was studied in Ref.
\cite{arXiv:1006.2406}. In this paper, we develop BTZ solutions to
the case of existence of electrodynamics. We take into account the
linear Maxwell field as well as some of the known nonlinear
electrodynamics models in gravity's rainbow. We also investigate
thermodynamical behavior of the solutions and discuss thermal
stability of them.

\section{BTZ and asymptotically BTZ Gravity's Rainbow}

The Lagrangian of Einstein gravity with cosmological constant coupled to an
electromagnetic field may be written as
\begin{equation}
L_{\mathrm{tot}}=R-2\Lambda -\mathcal{L}(\mathcal{F}),  \label{Lagrangian}
\end{equation}
where $R$ and $\Lambda $ are, respectively, the Ricci scalar and the
cosmological constant, and electromagnetic Lagrangian, $\mathcal{L}(\mathcal{%
F})$, is a function of Maxwell invariant $\mathcal{F}=F_{ab}F^{ab}$, where
the Faraday tensor is $F_{ab}=\partial _{\lbrack a}A_{b]}$ and $A_{b}$ is
the gauge potential. Taking into account the gauge-gravity Lagrangian (\ref%
{Lagrangian}) and applying the variational method, we obtain
\begin{equation}
G_{ab}+\Lambda g_{ab}=T_{ab},  \label{FE1}
\end{equation}
\begin{equation}
\nabla _{a}\left( \mathcal{L}_{\mathcal{F}}F^{ab}\right) =0,  \label{FE2}
\end{equation}%
where $G_{ab}$ is the Einstein tensor, $\mathcal{L}_{\mathcal{F}}=\frac{d%
\mathcal{L}(\mathcal{F})}{d\mathcal{F}}$ and
\begin{equation}
T_{ab}=\frac{1}{2}g_{\mu \nu }L(\mathcal{F})-2F_{\mu \lambda }F_{\nu
}^{\;\lambda }L_{\mathcal{F}},  \label{EM}
\end{equation}

In this paper, in addition to the linear Maxwell electrodynamics
(MED), we consider four classes of NED fields, namely BI nonlinear
electrodynamics (BINED), Exponential form of nonlinear
electrodynamics (ENED), Logarithmic form of nonlinear
electrodynamics (LNED) and Power-Maxwell invariant form of
electrodynamics (PNED) in which their Lagrangians are
\cite{BI,Soleng,HendiJHEP,HendiAnn1,HendiAnn2,PMIA1,PMIA2,PMIA3,PMIB1,PMIB2,PMIB3,PMIB4}
\begin{equation}
L(\mathcal{F})=\left\{
\begin{array}{cc}
-\mathcal{F}, & \;\;{MED} \vspace{0.3cm} \\
4\beta ^{2}\left( 1-\sqrt{1+\frac{\mathcal{F}}{2\beta ^{2}}}\right) , & \;\;{%
BINED} \vspace{0.3cm} \\
\beta ^{2}\left( \exp (-\frac{\mathcal{F}}{\beta ^{2}})-1\right) , & \;\;{%
ENED} \vspace{0.3cm} \\
-8\beta ^{2}\ln \left( 1+\frac{\mathcal{F}}{8\beta ^{2}}\right) , & \;\;{LNED%
} \vspace{0.3cm} \\
(-\mathcal{F})^{\alpha }, & \;\;{PNED}%
\end{array}%
\right. .  \label{LagEM}
\end{equation}%
In these equations, $\beta $ and $\alpha $ are called the nonlinearity
parameters. It is notable that for $\beta \longrightarrow \infty $ and $%
\alpha \longrightarrow 1$, all the mentioned nonlinear models reduce to
Maxwell Lagrangian.


Before we proceed, we would like to provide some reasonable
motivations for considering these forms of NED.

Although Maxwell theory is capable for describing different
phenomena in classical electrodynamics, it is not flawless and has
some fundamental problems \cite{Delphenich1,Delphenich2}. In order
to solve these
problems, we can generalize Maxwell theory to the case of NED \cite%
{HendiJHEP,HendiAnn1,HendiAnn2,PMIA1,PMIA2,PMIA3,PMIB1,PMIB2,PMIB3,PMIB4,Delphenich1,Delphenich2,Altshuler}.
Another strong motivation of considering NED, comes from
developments in string/M-theory. It has been shown that the
Born--Infeld \cite{BI} (BI)--type theories are specific in the
context of NED models, which are naturally acquired in the
low--energy
limit of heterotic string theory \cite%
{BIString1,BIString2,BIString3,BIString4}. In addition, it is easy
to show that the weak field limit of BI--type models is the same
as calculation of the one-loop approximation of QED, which leads
to quadratic Maxwell invariant in addition to the Maxwell
Lagrangian \cite{QED}.

Recently two BI--types of NED have been introduced, in which we
can remove the divergency of the electric field of point-like
charge near the origin. It was shown that BI--types of NED enjoy
special and important properties. To name a few, one can indicate
the absence of shock waves, birefringence \cite{Boillat1,Boillat2}
and an electric---magnetic duality \cite{Gibbons}. It was shown
that in a uniform electromagnetic field background for electrons,
calculation of exact one--loop corrections leads to presence of
logarithmic form of electromagnetic field strength \cite{HE}. This
correction term comes from the vacuum polarization effects. On the
other hand, BI--type theories have been employed for explaining
the equation of state of radiation for inflation \cite{Altshuler}.
In addition, in the context of the holographic superconductors, it
was shown that these NED models make significant effects on
condensation, critical temperature of the superconductor and its
energy gap [see \cite{Jing1,Jing2,Jing3,Dey1,Dey2} for more
details]. It was shown that the exponential form of NED has
stronger effect on the condensation formation and conductivity
comparing to logarithmic form of NED \cite{Dey1,Dey2}.

Now let us present some motivations of considering PMI theory. The
PMI theory is significantly richer than the Maxwell field, and in
the
special case ($s=1$) it reduces to linear electrodynamics \cite%
{PMIB1,PMIB2,PMIB3,PMIB4}. One of the most important properties of
the PMI model in $d$-dimensions occurs for $s=d/4$ where the PMI
theory becomes conformally invariant and so the energy-momentum
tensor will be traceless (the same as Maxwell theory in
four-dimensions \cite{PMIA1,PMIA2,PMIA3}). In this case, one can
obtain an inverse square law for the electric field of point-like
charge in arbitrary dimensions. Furthermore, it was shown that
there is an interesting relation between the black hole solutions
of a class of pure $F(R)$ gravity and those of conformally
invariant Maxwell source in Einstein gravity \cite{HendiFR}. In
the context of AdS/CFT correspondence, the effects of PMI source
on strongly coupled dual gauge theory have been investigated
\cite{Pan} (for more motivations of PMI theory, we refer the
reader to \cite{HendiVahidi}). Motivated by the recent results
mentioned above, we consider the linear and nonlinear
electrodynamics in three dimensional gravity's rainbow.

It was shown that adding the curvature to doubly special relativity (DSR)
\cite{DSR} leads to an extension, called doubly general relativity (DGR)
\cite{Smolin}. In DGR the spacetime geometry depends on the energy of the
particle ($E$). Thus, one may regard a spacetime which is parameterized by
the energy ratio ${\varepsilon }$ to form a rainbow of metrics. Three
dimensional static metric in the framework of gravity's rainbow is given by
\cite{Smolin}
\begin{equation}
d\tau ^{2}=-ds^{2}=\frac{\Psi (r)}{f^{2}({\varepsilon })}dt^{2}-\frac{1}{%
g^{2}({\varepsilon })}\left( \frac{dr^{2}}{\Psi (r)}+{r^{2}}d\phi
^{2}\right) .  \label{Metric}
\end{equation}

In this paper, we are looking for the black hole solutions with a radial
electric field. Thus the nonzero components of the gauge potential is
temporal component. Regarding the mentioned models of Eq. (\ref{LagEM}) and
using the gauge potential $A_{\mu }=h(r)\delta _{\mu }^{0}$, we find that
the electromagnetic field equation (\ref{FE2}) leads to the following
differential equations%
\begin{equation}
\begin{array}{ll}
rh^{\prime \prime }+h^{\prime }=0, & \;\;{MED} \\
&  \\
rh^{\prime \prime }+h^{\prime }\left[ 1-\left( \frac{f({\varepsilon })g({%
\varepsilon })h^{\prime }}{\beta }\right) ^{2}\right] =0, & \;\;{BINED} \\
&  \\
r\left[ 1+\left( \frac{2f({\varepsilon })g({\varepsilon })h^{\prime }}{\beta
}\right) ^{2}\right] h^{\prime \prime }+h^{\prime }=0, & \;\;{ENED} \\
&  \\
\begin{array}{l}
rh^{\prime \prime }\left[ 1+\left( \frac{f({\varepsilon })g({\varepsilon }%
)h^{\prime }}{2\beta }\right) ^{2}\right] + \\
\left[ 1-\left( \frac{f({\varepsilon })g({\varepsilon })h^{\prime }}{2\beta }%
\right) ^{2}\right] h^{\prime }=0,%
\end{array}
& \;\;{LNED} \\
&  \\
r\left( 2\alpha -1\right) h^{\prime \prime }+h^{\prime }=0,\;\;{\ \ \ \ }%
(\alpha \neq 1) & \;\;{PNED}%
\end{array}%
,  \label{heq}
\end{equation}%
where prime and double primes denote first and second derivative with
respect to $r$, respectively. The solutions of the mentioned differential
equations are
\begin{equation}
h(r)=\left\{
\begin{array}{cc}
q\ln \left( \frac{r}{l}\right) , & \;\;{MED} \\
&  \\
q\ln \left[ \frac{r}{2l}\left( 1+\Gamma \right) \right] , & \;\;{BINED} \\
&  \\
\multicolumn{1}{l}{%
\begin{array}{c}
q\left[ \exp \left( -\frac{L_{W}}{2}\right) +\frac{E_{i}\left( 1,\frac{L_{W}%
}{2}\right) }{2}+\right. \\
\left. \ln \left( \frac{\sqrt{2}f({\varepsilon })g({\varepsilon })q}{\beta l}%
\right) +\frac{\gamma }{2}-1\right]%
\end{array}%
,} & \;\;{ENED} \\
&  \\
\multicolumn{1}{l}{\frac{\beta ^{2}r^{2}\left( \Gamma -1\right) }{qf({%
\varepsilon })^{2}g({\varepsilon })^{2}}-\frac{q}{2}+q\ln \left[ \frac{%
r\left( 1+\Gamma \right) }{2l}\right] ,} & \;\;{LNED} \\
&  \\
\frac{\left( 2s-1\right) q}{2(s-1)r^{\frac{2(1-s)}{2s-1}}},\;\;{\ }(s\neq 1)
& \;\;{PNED}%
\end{array}%
\right. ,  \label{h(r)}
\end{equation}%
where $\Gamma =\sqrt{1+\frac{f^{2}({\varepsilon })g^{2}({\varepsilon })q^{2}%
}{r^{2}\beta ^{2}}}$, $q$ is an integration constant which is related to the
electric charge of the black hole. In addition, $L_{W}=LambertW\left( \frac{%
4f^{2}({\varepsilon })g^{2}({\varepsilon })q^{2}}{\beta ^{2}r^{2}}\right) $\
which satisfies $LambertW(x)\exp \left[ LambertW(x)\right] =x$, $\gamma
=\gamma (0)\simeq 0.57722$\ and the special function $Ei\left( 1,x\right)
=\int\limits_{1}^{\infty }\frac{e^{-xz}}{z}dz$\ (for more details, see \cite%
{LambertW}).

According to the gauge potential ansatz with Eq. (\ref{h(r)}), one finds
that the non-vanishing components of the electromagnetic field tensor can be
written as
\begin{equation}
F_{tr}=-F_{rt}=\left\{
\begin{array}{cc}
\frac{q}{r}, & \;\;{MED} \\
&  \\
\frac{q}{r\Gamma }, & \;\;{BINED} \\
&  \\
\multicolumn{1}{l}{\frac{q}{r}\exp \left( -\frac{L_{W}}{2}\right) ,} & \;\;{%
ENED} \\
&  \\
\multicolumn{1}{l}{\frac{2r\beta ^{2}}{qf^{2}({\varepsilon })g^{2}({%
\varepsilon })}\left( \Gamma -1\right) ,} & \;\;{LNED} \\
&  \\
\frac{q}{r^{\frac{1}{2\alpha -1}}}, & \;\;{PNED}%
\end{array}%
\right. .  \label{Ftr}
\end{equation}%
In order to have a sensible asymptotic structure for the electromagnetic
field, $F_{tr}$ should vanish for large values of $r$. This condition is
satisfied for MED, BINED, ENED and LNED branches. For PNED case, vanishing $%
F_{tr}$ for $r\rightarrow \infty $ leads to $\alpha >1/2$. Thus in this
paper, we restrict ourselves to $\alpha >1/2$. Regarding the gravitational
field equation for the mentioned models of NED, we find that the following
metric function satisfies all components of the field equation (\ref{FE1}),
simultaneously
\begin{equation}
\Psi (r)=-\frac{\Lambda r^{2}}{g(E)^{2}}-m+\Theta (r),  \label{Psi}
\end{equation}%
where
\begin{equation}
\Theta (r)=\left\{
\begin{array}{cc}
-2q^{2}f^{2}({\varepsilon })\ln \left( \frac{r}{l}\right) , & \;\;{MED}%
\vspace{0.3cm} \\
&  \\
\begin{array}{c}
\frac{2r^{2}\beta ^{2}\left( 1-\Gamma \right) }{g^{2}({\varepsilon })}%
+q^{2}f^{2}({\varepsilon })-\vspace{0.3cm} \\
2q^{2}f^{2}({\varepsilon })\ln \left( \frac{r\left( 1+\Gamma \right) }{2l}%
\right) ,%
\end{array}
& \;\;{BINED}\vspace{0.3cm} \\
&  \\
\begin{array}{c}
q^{2}f^{2}({\varepsilon })E_{i}\left( 1,\frac{L_{W}}{2}\right) +\frac{\beta
^{2}r^{2}}{2g^{2}({\varepsilon })}+\vspace{0.2cm} \\
\frac{q^{2}f^{2}({\varepsilon })\left( \gamma -3+\ln 2\right) }{g^{2}({%
\varepsilon })}-\frac{q\beta rf({\varepsilon })}{g({\varepsilon })\sqrt{LW}}+%
\vspace{0.2cm} \\
2q^{2}f^{2}({\varepsilon })\ln \left( \frac{f({\varepsilon })g({\varepsilon }%
)q}{\beta l}\right) +\vspace{0.2cm} \\
4q^{2}f^{2}({\varepsilon })\exp \left( \frac{-L_{W}}{2}\right) ,%
\end{array}
& {ENED}\vspace{0.3cm} \\
&  \\
\begin{array}{c}
f^{2}({\varepsilon })q^{2}\left\{ -\ln \left[ \frac{\beta ^{2}r^{4}\left(
\Gamma -1\right) \left( \Gamma +1\right) ^{3}}{4q^{2}l^{2}f^{2}({\varepsilon
})g^{2}({\varepsilon })}\right] +\right. \vspace{0.2cm} \\
\left. \frac{4\beta ^{2}r^{2}}{f^{2}({\varepsilon })g^{2}({\varepsilon }%
)q^{2}}\left[ \ln \left( \frac{\Gamma +1}{2}\right) +\frac{3}{2}\right]
-\right. \vspace{0.2cm} \\
\left. 6\beta ^{2}r^{2}\Gamma +2\right\}%
\end{array}%
, & {LNED}\vspace{0.3cm} \\
&  \\
-\frac{2^{\alpha }\left[ f^{2}({\varepsilon })g^{2}({\varepsilon })q^{2}%
\right] ^{\alpha }\left( 2\alpha -1\right) ^{2}}{2(\alpha -1)g^{2}({%
\varepsilon })r^{\frac{2(1-\alpha )}{2\alpha -1}}}, & \;\;{PNED}%
\end{array}%
\right. ,  \label{Theta}
\end{equation}%
and $m$ is an integration constant which is related to the total
mass ($M$)
\begin{equation}
M=\frac{m}{8f({\varepsilon })}.  \label{Mass}
\end{equation}

It is notable that the Arnowitt–Deser–Misner (ADM) mass of black
hole can be obtained by using the behavior of the metric at large
$r$. Straightforward calculations show that ADM mass of the black
hole solutions is the same as Eq. (\ref{Mass}).

Using series expansion of metric functions for large values of $r$, we find
that $\Lambda $-term will be dominated and therefore obtained solutions are
asymptotically AdS with an effective cosmological constant, $\Lambda _{eff}=%
\frac{\Lambda }{g(E)^{2}}$. In addition, calculations show that the Ricci
and the Kretschmann scalars for the metric (\ref{Metric}) are
\begin{eqnarray}
R &=&-g^{2}({\varepsilon })\Psi ^{\prime \prime }-\frac{2g^{2}({\varepsilon }%
)\Psi ^{\prime }}{r}  \label{R} \\
R_{\mu \nu \rho \sigma }R^{\mu \nu \rho \sigma } &=&g^{4}({\varepsilon }%
)\Psi ^{\prime \prime 2}+2\left( \frac{g^{2}({\varepsilon })\Psi ^{\prime }}{%
r}\right) ^{2}.  \label{RR}
\end{eqnarray}

Inserting Eq. (\ref{Psi}) into the Eqs. (\ref{R}) and (\ref{RR}),
we find that there is an essential singularity located at $r=0$.
Depending on the values of free parameters, one confirms that the
mentioned singularity can be covered with an event horizon
\cite{HendiJHEP}. Applying the surface gravity interpretation on
the event horizon, one finds the Hawking temperature as
\begin{equation}
T=-\frac{\Lambda r_{+}}{2\pi f({\varepsilon })g({\varepsilon })}+\Upsilon
_{+},  \label{T}
\end{equation}
where
\begin{equation}
\Upsilon _{+}=\left\{
\begin{array}{cc}
\frac{-f({\varepsilon })g({\varepsilon })q^{2}}{2\pi r_{+}}, & \;\;{MED} \\
&  \\
\frac{r\beta ^{2}\left( 1-\Gamma _{+}\right) }{\pi f({\varepsilon })g({%
\varepsilon })}, & \;\;{BINED} \\
&  \\
\frac{q\beta \left( 1-L_{W+}\right) }{2\pi \sqrt{L_{W+}}}-\frac{\beta
^{2}r_{+}}{4\pi f({\varepsilon })g({\varepsilon })}, & \;\;{ENED} \\
&  \\
\frac{2\beta ^{2}r_{+}\left[ 1+\ln \left( \frac{1+\Gamma _{+}}{2}\right)
-\Gamma _{+}\right] }{\pi f({\varepsilon })g({\varepsilon })}, & \;\;{LNED}
\\
&  \\
\frac{-\left[ 2f^{2}({\varepsilon })g^{2}({\varepsilon })q^{2}\right]
^{\alpha }(2\alpha -1)}{4\pi f({\varepsilon })g({\varepsilon })r_{+}^{\frac{1%
}{2\alpha -1}}}, & \;\;{PNED}%
\end{array}%
\right. ,
\end{equation}%
and $\Gamma _{+}=\sqrt{1+\frac{f^{2}({\varepsilon })g^{2}({\varepsilon }%
)q^{2}}{r_{+}^{2}\beta ^{2}}}$ and $L_{W_{+}}=LambertW\left( \frac{4f^{2}({%
\varepsilon })g^{2}({\varepsilon })q^{2}}{\beta ^{2}r_{+}^{2}}\right) $.

It is worthwhile to mention that in the absence of rainbow's effects ($f({%
\varepsilon })=g({\varepsilon })=1$), there is a critical value for the
horizon radius ($r_{+c}$), in which the temperature is positive for $%
r_{+}>r_{+c}$. The value of $r_{+c}$ depends on the choices of other
parameters. In order to have a consistent limitation, we restrict $%
f(\varepsilon )$ and $g(\varepsilon )$ to smooth positive functions. It is
evident that due to the presence of rainbow's functions, the behavior of
temperature and the values of $r_{+c}$ change, drastically.

\section{First law of thermodynamics and thermal stability \label{Thermo}}

In order to check the First law of thermodynamics, we should
calculate entropy, electric potential and charge. Since we are
working in Einstein gravity, it is allowed to use the area law to
obtain the entropy \cite{Area1,Area2,Area3}. The entropy of black
holes is equal to one-quarter of the horizon area
\begin{equation}
S=\frac{\pi r_{+}}{2g(\varepsilon )}.  \label{S}
\end{equation}

Besides, we can obtain the electric charge by using the Gauss law.
By calculation of the flux of the electric field at infinity, one
can find the electric charge as
\begin{equation}
Q=\left\{
\begin{array}{cc}
\frac{qf(\varepsilon )}{2}, &
\begin{array}{c}
\;\;{MED} \\
{BINED} \\
{ENED} \\
{LNED}%
\end{array}
\\
\begin{array}{c}
\\
\frac{2^{\alpha -2}\alpha }{g(\varepsilon )}\left[ qf(\varepsilon
)g(\varepsilon )\right] ^{2\alpha -1},%
\end{array}
&
\begin{array}{c}
\;\; \\
{PNED}%
\end{array}%
\end{array}%
\right. ,  \label{Q}
\end{equation}%
where for $\alpha =1$, the electric charge of PNED branch is compatible with
other branches. The electric potential $\Phi $, measured at the reference
with respect to the horizon, is defined by
\begin{equation}
\Phi =A_{\mu }\chi ^{\mu }\left\vert _{reference}-A_{\mu }\chi ^{\mu
}\right\vert _{r=r_{+}},  \label{Phi}
\end{equation}%
where $\Phi $ vanishes at the reference and $\chi ^{\mu }$ is the null
generator of the horizon, yielding
\begin{equation}
\Phi =\left\{
\begin{array}{cc}
-q\ln \left( \frac{r_{+}}{l}\right) , & \;\;{MED} \\
&  \\
-q\ln \left[ \frac{r_{+}}{2l}\left( 1+\Gamma _{+}\right) \right] , & \;\;{%
BINED} \\
&  \\
\multicolumn{1}{l}{%
\begin{array}{l}
q\left[ 1-\exp \left( -\frac{L_{W_{+}}}{2}\right) -\frac{\gamma }{2}-\right.
\\
\left. \ln \left( \frac{\sqrt{2}f(\varepsilon )g(\varepsilon )q}{\beta l}%
\right) -\frac{E_{i}\left( 1,\frac{L_{W_{+}}}{2}\right) }{2}\right] ,%
\end{array}%
} & \;\;{ENED} \\
&  \\
\multicolumn{1}{l}{\frac{q}{2}+\frac{\beta ^{2}r_{+}^{2}\left( 1-\Gamma
_{+}\right) }{qf^{2}(\varepsilon )g^{2}(\varepsilon )}-q\ln \left[ \frac{%
\left( 1+\Gamma _{+}\right) r_{+}}{2l}\right] ,} & \;\;{LNED} \\
&  \\
\frac{-\left( 2\alpha -1\right) q}{2(\alpha -1)r_{+}^{\frac{2(1-\alpha )}{%
2\alpha -1}}},(\alpha \neq 1) & \;\;{PNED}%
\end{array}%
\right.  \label{U}
\end{equation}

After calculating all of the conserved and thermodynamic
quantities of the solutions, we are in a position to examine the
first law of thermodynamics. To do this, we obtain the total mass
$M$ as a function of the extensive quantities $Q$ and $S$. Using
the expression for the entropy, the electric charge and the mass
given in Eqs. (\ref{Psi}), (\ref{Mass}), (\ref{S}) and (\ref{Q}),
and the fact that $f(r=r_{+})=0$, one can obtain a Smarr-type
formula as
\begin{equation}
M\left( S,Q\right) =-\frac{\Lambda S^{2}}{2\pi ^{2}f(\varepsilon )}-\frac{%
\Sigma }{f(\varepsilon )},  \label{Msmar}
\end{equation}%
where%
\begin{equation}
\Sigma =\left\{
\begin{array}{cc}
Q^{2}\ln \left( \frac{2g(\varepsilon )S}{\pi l}\right) , & \;\;{MED}\vspace{%
0.3cm} \\
&  \\
Q^{2}\ln \left( \frac{g(\varepsilon )(1+\Gamma ^{\prime })S}{\pi l}\right)
-2\left( \frac{\beta (1-\Gamma ^{\prime })S}{2\pi }\right) ^{2}, & \;\;{BINED%
}\vspace{0.2cm} \\
&  \\
\begin{array}{c}
\frac{Q^{2}}{2}\left[ Ei\left( 1,\frac{L_{W}^{\prime }}{2}\right) +\gamma -3%
\right] +\vspace{0.2cm} \\
\frac{Q^{2}}{2}\ln \left( \frac{8Q^{2}g^{2}(\varepsilon )}{\beta ^{2}l^{2}}%
\right) +\frac{\beta ^{2}S^{2}}{4\pi ^{2}}-\vspace{0.2cm} \\
\frac{\beta QS\left( 1-2L_{W}^{\prime }\right) }{2\pi \sqrt{L_{W}^{\prime }}}%
,%
\end{array}
& {ENED} \\
&  \\
\begin{array}{c}
\frac{Q^{2}}{8}\left[ \frac{(3-\Gamma ^{\prime 2})\ln \left( \frac{1+\Gamma
^{\prime }}{2}\right) }{(1-\Gamma ^{\prime 2})}+\frac{(2-\Gamma ^{\prime })}{%
(1+\Gamma ^{\prime })}+\right. \\
\left. \ln \left( \frac{2g(\varepsilon )S}{\pi l}\right) \right] ,%
\end{array}
& \;\;{LNED} \\
&  \\
\frac{(2\alpha -1)^{2}\left( \frac{2^{\alpha }Q}{\alpha }\right) ^{\frac{%
2\alpha }{2\alpha -1}}}{4(\alpha -1)\left( \frac{S}{\pi }\right) ^{\frac{%
2-2\alpha }{2\alpha -1}}}, & \;\;{PNED}%
\end{array}%
\right.
\end{equation}%
and $\Gamma ^{\prime }=\sqrt{1+\frac{\pi ^{2}Q^{2}}{\beta ^{2}S^{2}}}$ and $%
L_{W}^{\prime }=LambertW\left( \frac{4\pi ^{2}Q^{2}}{\beta ^{2}S^{2}}\right)
$. Taking into account Eq. (\ref{Msmar}), we can calculate the temperature
and electric potential as intensive parameters
\begin{equation}
T=\left( \frac{\partial M}{\partial S}\right) _{Q}\ ,\ \ \ \ \Phi =\left(
\frac{\partial M}{\partial Q}\right) _{S},\   \label{TPhi}
\end{equation}%
which are the same as those calculated in Eqs. (\ref{T}) and (\ref{Phi})
(for all branches). Therefore we conclude that obtained conserved and
thermodynamic quantities satisfy the first law of thermodynamics
\begin{equation}
dM=TdS+\Phi dQ.  \label{Firstk1}
\end{equation}

In what follows, we investigate thermal stability. In principle,
thermal stability of a system with respect to small variations of
the thermodynamic coordinates is usually performed by
investigating the behavior of the entropy $S(M,Q)$ around the
equilibrium.

The local stability criteria in any ensemble requires that
$S(M,Q)$ be a smooth convex function of the extensive variables or
its Legendre transformation should be a smooth concave function of
the intensive thermodynamic coordinates. Thermal stability can
also be studied by the behavior of the total mass $M(S,Q)$ which
should be a smooth convex function of its extensive variables.
Therefore, the local stability can in principle be carried out by
obtaining the determinant of the Hessian matrix of $M(S,Q)$ versus
its extensive coordinates $X_{i}$,
$\mathbf{H}_{X_{i}X_{j}}^{M}=[\partial ^{2}M/\partial
X_{i}\partial X_{j}]$ (see Refs.
\cite{Stability1,Stability2,Stability3} for more details). In our
case the mass $M$ is a function of the entropy $S$ and the charge
$Q$. The number of thermodynamic variables depends on the ensemble
that is used. In what follows, we calculate the heat capacity of
(nonlinearly) charged black hole solutions to investigate thermal
stability through canonical ensemble. We should regard the
electric charge as a fixed parameter and calculate $C_{Q}=T\left(
\frac{\partial S}{\partial T}\right) _{Q}$. The positivity of the
heat capacity guarantees the local thermal stability. Since we
investigate the physical black hole solutions (positive
temperature), it is sufficient to examine the positivity of
$\left( \frac{\partial S}{\partial T}\right) _{Q}=\left(
\frac{\partial ^{2}M}{\partial S^{2}}\right) _{Q}^{-1}$ for
thermal stability. Calculations show that
\begin{equation}
\left( \frac{\partial ^{2}M}{\partial S^{2}}\right) _{Q}=\frac{-\Lambda }{%
\pi ^{2}f(\varepsilon )}+\left\{
\begin{array}{cc}
\frac{q^{2}f(\varepsilon )g^{2}(\varepsilon )}{\pi ^{2}r_{+}^{2}}, & \;\;{MED%
} \\
&  \\
\frac{2\beta ^{2}\left( \Gamma _{+}-1\right) }{\pi ^{2}f(\varepsilon )\Gamma
_{+}}, & \;\;{BINED} \\
&  \\
\frac{\beta ^{2}\left( e^{\frac{L_{W+}}{2}}-1\right) }{2\pi
^{2}f(\varepsilon )}, & \;\;{ENED} \\
&  \\
\frac{4\beta ^{2}\ln \left( \frac{1+\Gamma _{+}}{2}\right) }{\pi
^{2}f(\varepsilon )}, & \;\;{LNED} \\
&  \\
\frac{2^{\alpha }\left[ qf(\varepsilon )g(\varepsilon )\right] ^{2\alpha }}{%
2\pi ^{2}f(\varepsilon )r_{+}^{\frac{2\alpha }{2\alpha -1}}}, & \;\;{PNED}%
\end{array}%
\right. ,  \label{dMdSS}
\end{equation}%
It is clear to find that $\Gamma _{+}$ and $e^{\frac{L_{W+}}{2}}$ are
greater than one and therefore $\left( \frac{\partial ^{2}M}{\partial S^{2}}%
\right) _{Q}$ is positive definite for obtained AdS solutions. Regarding
positive values for $f(\varepsilon )$ and $g(\varepsilon )$ and taking into
account a positive temperature for the physical black hole solutions ( for $%
r_{+}>r_{+c}$), we can conclude that the mentioned black holes
enjoy thermal stability. In comparison with the asymptotically AdS
black holes of Einstein gravity, which have a small unstable
phase, the stability phase structure of the Einstein-(nonlinear)
Maxwell black holes in gravity's rainbow shows that the electric
charge help us to obtain stable solutions and rainbow functions do
not make anomaly.

\section{Note on the integration constants}

Regarding positive real rainbow functions with the functional forms of $%
\Gamma $ and $L_{W}$ (after Eq. (\ref{h(r)})), one finds that
there is no restriction on $q$\ and $m$. Besides, one can follow
the method of Refs. \cite{NEW1,NEW2,NEW3} with a specific model of
rainbow function to obtain possible mass/charge limitations. Since
in this paper we have discussed general form of rainbow functions,
we release the mentioned method.

Taking into account Eqs. (\ref{h(r)}( and (\ref{Psi}), one can
find that $q$ and $m$ are integration constant. Since the variable
of integrations is $r$
coordinate, we can replace $q$ and $m$ with modified constants such as $%
qf^{a}(\varepsilon )g^{b}(\varepsilon )$ and $mf^{c}(\varepsilon
)g^{d}(\varepsilon )$, respectively ($a$, $b$, $c$ and $d$ are arbitrary
real valued numbers). It is worthwhile to mention that although in new
situation, all conserved and thermodynamic quantities will be modified, the
results for $\varepsilon \rightarrow 0$ ($f(\varepsilon )=g(\varepsilon )=1$%
) do not change. Another comment which should be discussed is the
first law and thermal stability. Regarding new situation (applying
$q\rightarrow qf^{a}(\varepsilon )g^{b}(\varepsilon )$ and
$m\rightarrow mf^{c}(\varepsilon )g^{d}(\varepsilon )$ in previous
sections), it is easy to show that the first law is satisfied. In
addition, the criteria of thermal stability do not change,
globally. These behaviors are expected, since we applied a global
transformation on the constants. Considering new situation, one
can obtain four free parameters $a$, $b$, $c$ and $d$, and their
manipulations with special choices may lead to interesting
results.

\section{Closing Remarks}

In this paper, we took into account the gravity's rainbow in three
dimensions in the presence of linear and nonlinear models of
electrodynamics. The motivation of considering gravity's rainbow
comes from the quantum gravity view point, while the reason of
considering nonlinear electrodynamics is arisen from LQG as well
as low effective limit of heterotic sting theory. We found that
although the rainbow's functions affected the electromagnetic
field behavior of nonlinear BI-type models, Maxwell and PNED
models of electromagnetic field did not depend on the rainbow's
functions. We obtained metric functions of each models and found
that these solutions can be interpreted as black holes.

Then, we calculated conserved and thermodynamic quantities and
found that these solutions may depend on the choice of rainbow's
functions. We also considered finite mass as a function of
extensive parameters and checked the first law of thermodynamics.
In addition, we analyzed thermal stability by using of the heat
capacity. We found that obtained physical solutions ($T>0$)
enjoyed thermal stability in the canonical ensemble.

It is worthwhile to generalize these three dimensional solutions to four and
higher dimensional cases with various horizon topologies. It is also
interesting to investigate the effects of rainbow's functions on dilaton
gravity as well as higher curvature theory. These cases will be addressed
elsewhere.

\section{acknowledgements}
We are indebted to Mir Faizal for useful discussions and S.
Panahiyan for reading the manuscript. We also wish to thank Shiraz
University Research Council. This work has been supported
financially by Research Institute for Astronomy and Astrophysics
of Maragha.

\end{document}